\documentclass[%
reprint,
amsmath,amssymb,
aps,
 prx,
floatfix,
twocolumn]{revtex4-2}

\usepackage{hyperref}

\usepackage{graphicx}
\usepackage{dcolumn}
\usepackage{bm}
\usepackage{amsmath}
\usepackage{amssymb}
\usepackage{mathrsfs}
\usepackage{siunitx}
\usepackage{xcolor}
\usepackage{tabularx}
\usepackage{microtype}
\usepackage{graphicx,wrapfig,lipsum}
\newcommand{\fig}{Fig.~}

\newcommand{\fs}[1]{\textbf{#1}} 

\begin{document}
\title{Finite integration time can shift optimal sensitivity away from criticality} 

\author{Sahel Azizpour$^{1,2}$}
\thanks{Current Affiliation: Donders Institute for Brain, Cognition, and Behaviour, Radboud University}
\author{Viola Priesemann$^{3,4}$}
\author{Johannes Zierenberg$^{3,4}$}
\thanks{JZ and AL contributed equally}
\author{Anna Levina$^{1,2}$}
\thanks{JZ and AL contributed equally}

\affiliation{
  $^1$\mbox{Department of Computer Science, University of Tübingen, T{\"u}bingen, Germany}
  $^2$\mbox{Max Planck Institute for Biological Cybernetics, T{\"u}bingen, Germany}
  $^3$\mbox{Max Planck Institute for Dynamics and Self-Organization, G{\"o}ttingen, Germany},\\
  $^4$\mbox{Institute for the Dynamics of Complex Systems, University of G\"ottingen, G\"ottingen, Germany},\\
}

\date{\today}

\begin{abstract}

Sensitivity to small changes in the environment is crucial for many real-world tasks, enabling living and artificial systems to make correct behavioral decisions.
It has been shown that such sensitivity is maximized when a system operates near the critical point of a phase transition.
However, proximity to criticality introduces large fluctuations and diverging timescales.
Hence, to leverage the maximal sensitivity, it would require impractically long integration periods.
Here, we analytically and computationally demonstrate how the optimal tuning of a recurrent neural network is determined given a finite integration time.
Rather than maximizing the theoretically available sensitivity, we find networks attain different sensitivities depending on the available time.
Consequently, the optimal dynamic regime can shift away from criticality when integration times are finite, highlighting the necessity of incorporating finite-time considerations into studies of information processing.
\end{abstract}

\maketitle

Living systems must efficiently encode relevant environmental information while being sensitive to small changes.
Increasing evidence suggests that many natural systems tackle this challenge by operating near a critical phase transition~\cite{munoz_colloquium_2018}.
Signatures of near-critical dynamics have been observed across different scales, from collective behaviors in flocks of birds~\cite{cavagna_scale-free_2010} to cellular diversity in stem cell populations~\cite{ridden_entropy_2015}, and most notably in the brain~\cite{beggs_criticality_2008, priesemann_subsampling_2009, palva_neuronal_2013, wilting_25_2019, fontenele_low-dimensional_2024}.
The proposed advantage of operating near a critical point is that phase transitions endow systems with computational benefits, including elevated sensitivity and correlation~\cite{henkel_non-equilibrium_2008, tauber_critical_2014}, maximized dynamic range~\cite{kinouchi_optimal_2006}, enhanced information flow~\cite{boedecker_information_2012, barnett_information_2013, meijers_behavior_2021}, optimal input representation~\cite{morales_optimal_2021, yang_critical_2025}, and a diverse spectrum of dynamical responses~\cite{nykter_critical_2008}.

Operating {in the vicinity of a} critical phase transition offers significant advantages but comes with inherent challenges.
While enhanced sensitivity of critical systems makes them ideal for some tasks, it also increases their vulnerability to noise, further amplified by critical slowing down~\cite{dakos_slowing_2008, maturana_critical_2020}.
A recent example of this is decision-making by integrated Ising models, where operating at a distance from a phase transition allows to control the trade-off between reaction time and error rate~\cite{tapinova_integrated_2025}.
More generally, such a trade-off can be formulated as an optimization problem with a control parameter $\lambda$ (in our case changing the distance to criticality) that regulates both beneficial gain $G(\lambda)$ and detrimental loss $L(\lambda)$ with some weighting factor $\gamma$, i.e.,
\begin{equation}
\lambda^\ast = \underset{\lambda}{\text{arg max}} \, \left\{G(\lambda) - \gamma L(\lambda)\right\}.
\end{equation}

Both gains and losses depend on the particularities of the system. Thus, the optimal tuning of $\lambda$, and thereby the optimal distance to criticality, will have to depend on the specific system and requirements of each task~\cite{wilting_operating_2018}.
For example, fish schools balance reaction time and energy cost in their alarmed state~\cite{poel_subcritical_2022}, while neuromorphic computing and artificial networks adjust their state to match memory requirements for optimal functioning~\cite{cramer_control_2020, khajehabdollahi_emergent_2024}.
Despite these observations, it remains a challenge to quantitatively assess the trade-off between gain and loss that would determine an optimal distance from criticality.

A famous example of how criticality can assist encoding in the brain is the dynamic range.
The dynamic range quantifies the range of continuous input features that can be encoded by the nonlinear firing-rate response of a neuron.
It is commonly defined as the logarithmic range of inputs $h$ for which the output is between 10$^{\mathrm{th}}$ and 90$^{\mathrm{th}}$ percentile of all outputs~\cite{kinouchi_optimal_2006}, i.e., $\Delta=10\log_{10}(h_{0.9}/h_{0.1})$, selected to exclude responses that would not be distinguishable from the noise floor at low activity and saturation regime at high activity.
Examples include encoding of correlations in the visual field~\cite{britten_analysis_1992}, odor concentration~\cite{wachowiak_representation_2001} and sound level~\cite{evans_dynamic_1981, dean_neural_2005}.

Unfortunately, the dynamic range of a single cell is usually much smaller than the dynamic range of perception.
This dynamic-range problem can be solved with the emergent properties from recurrent interactions, which were shown to drastically increase the dynamic range as the network approaches criticality~\cite{kinouchi_optimal_2006, shew_neuronal_2009, gautam_maximizing_2015}.
Exploiting close-to-critical emergence was also observed in structures with heterogeneous~\cite{gollo_coexistence_2017}, modular~\cite{zierenberg_tailored_2020}, or hierarchical~\cite{galera_physics_2020} organization.
However, previous work neglected the emerging close-to-criticality population activity fluctuations that can hinder confidence in discrimination.

\begin{figure*}[t]
    \centering
    \includegraphics[]{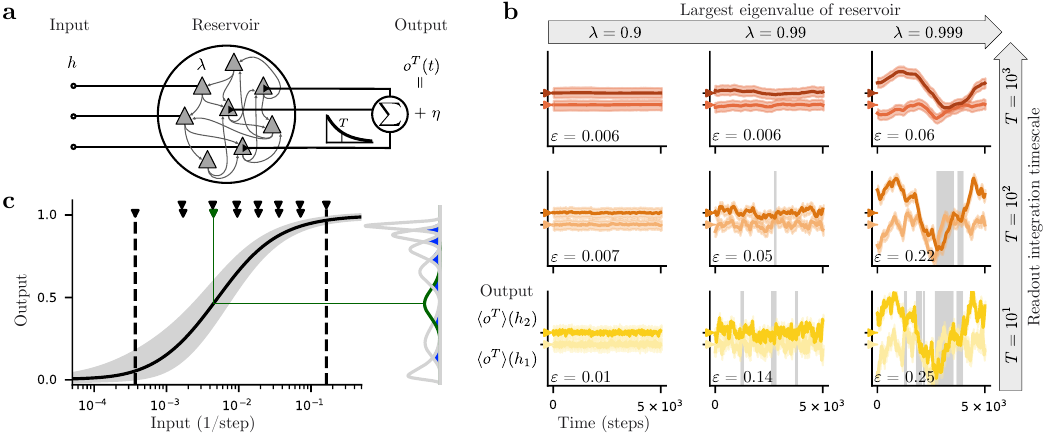}
    \caption{
    \textbf{Fluctuations in network activity can lead to unreliable input reconstruction.}
    \textbf{a)} Illustration of the neural reservoir:
    A subset of recurrently connected neurons with the largest eigenvalue of the connectivity matrix equal $\lambda$ receives Poisson spikes with rate $h$; the output integrates spikes of a subset of neurons with a timescale $T$ and is subject to noise $\eta$.
    %
    \textbf{b)}~Temporal evolution of output for different $\lambda$ and $T$ comparing the responses to two different inputs $h_1 < h_2$ with mean responses (marked by the arrows on the $y$-axis) fixed for all panels at the same values with $\langle o^T\rangle(h_1) < \langle o^T\rangle(h_2)$.
    Solid lines are the pure output response of the network, and opaque areas include noise.
    Gray regions highlight the times when the output responses differ from the order of their mean responses.
    The minimal discrimination error of the two underlying distributions is given by $\varepsilon$, cf. Eq.~\eqref{Eq:MDE}.
    \textbf{c)} Response curve indicating output values to logarithmic input rates.
    The solid black line shows the mean response, and the gray shading indicates the strength of fluctuations stemming from neural activity and output noise.
    Inputs are called discriminable when their output distributions have a sufficiently small overlap (see blue areas on the right).
    The first inputs that are discriminable from zero and full activity mark the dynamic range (black dashed lines).
    From these, we can construct sets of discriminable inputs marked by the black triangles (see text for details).
    }
    \label{fig1}
\end{figure*}

In this work, we combine analytical calculations, numerical simulations, and machine-learning approximations to quantify the optimal balance between input discrimination confidence and the sensitivity of a recurrent neural network, controlled by its recurrent interaction strength $\lambda$ and the timescale $T$ of a leaky readout (Fig.~\ref{fig1}).
To formalize this optimization problem, we introduce two generalized measures of dynamic range derived from the discriminability of inputs and provide analytical results for the limiting cases of instantaneous readout and infinite integration time.
We find that the optimal state, $\lambda^\ast$, of the network depends on the required confidence and integration time, with a safety margin from the precise critical point for all finite integration times.

We consider a random network of probabilistic spiking neurons that can be activated externally and recurrently (see Methods for details).
To mimic processing and transmission, only a random subset of neurons receives input, while another random subset of neurons serves as output (Fig.~\ref{fig1}a).
Input neurons receive uncorrelated, independent Poisson spike trains with a rate $h$, which represents the input strength.
The recurrent interactions are defined by a random sparse connectivity matrix $W$ for which we can control the largest eigenvalue $\lambda$ and thereby the fluctuations of recurrent activity.
A leaky readout integrates output neurons' activity with timescale $T$, which can be expressed as a sum of exponential kernels
\begin{equation}\label{eq:output}
    o^T(t) = \frac{1}{N^\mathrm{out}}\sum_{i\in N^\mathrm{out}}\sum_{\{k | \, t_i^k < t\}} e^{-(t-t_i^k)/T} + \eta,
\end{equation}
where $t_i^k$ is the timing of $k$-th spike of neuron $i$.
Here, we added a small Gaussian noise $\eta \sim \mathcal{N}(0,\sigma^2)$ to be able to technically treat $\delta$-distributed outputs from absorbing states or mean-field solutions in our later analysis with only a minor effect on typical output distributions (cf. Fig.~\ref{fig:extData_workflow}).
Depending on the parameters of the recurrent interactions, the input ordering would be more or less observable from the output activity, for an external observer or for neurons further up in the processing hierarchy.

For the extreme cases of $T \to \infty$ or $T \to 0$, we can solve the model analytically and obtain closed-form solutions for $P(o^T|h)$.
For $T \to \infty$ this can be achieved using a mean-field approach, and for $T \to 0$ we solve a Fokker-Plank equation (see Methods).
For intermediate integration times, $0<T<\infty$, we have to rely on simulations to obtain the output distributions.
This intermediate regime is particularly relevant for biological systems, as the intrinsic timescales of the cortical neurons were found to be in the range of about \SI{50}{} -- \SI{500}{ms}~\cite{murray_hierarchy_2014, rudelt_signatures_2024, shi_brain-wide_2025}. Comparable timescales can also arise from the slow accumulation of activity signals that inform or trigger synaptic plasticity, for instance, through calcium decay following neural spiking, typically lasting around \SI{500}{ms}~\cite{vogelstein_spike_2009}.

The sensitivity of the system is controlled by the largest eigenvalue $\lambda$ of the connectivity matrix~\cite{larremore_predicting_2011}.
The recurrent network we consider has a critical non-equilibrium phase transition for $h\to 0$ at $\lambda_c=1$~\cite{larremore_predicting_2011, zierenberg_tailored_2020}, where both sensitivity and correlated fluctuations are strongest.
Reducing $\lambda$ reduces recurrent network fluctuations (Fig.~\ref{fig1}b).
This generates a trade-off: close to criticality, we expect optimized information processing properties for infinite integration time but simultaneously increased fluctuations in the finite-time output $o^T(t)$.
In the following, we explore how to quantify this trade-off depending on the available integration time.

As a first step, we illustrate how the recurrent network's largest eigenvalue and the readout integration time affect the representation of inputs (\fig\ref{fig1}b).
Each panel shows the outputs from two identical copies of the network with given $\lambda$ and $T$ in response to the two stimuli with rates $h_1 < h_2$, which are chosen such that the mean-field outputs $\langle o^T\rangle (h_{2/1})=0.5\pm \Delta o/2$ are easily distinguishable with $\langle o^T \rangle(h_{2}) - \langle o^T \rangle(h_{1}) = \Delta o = 5\sigma$.
Assume the observer's task is to find which network received the stronger input.
This can be solved by deciding that a stronger output indicates a stronger input.
We shade gray the times where, following this strategy, the observer will make a mistake.
We observe only small fluctuations when $\lambda$ is far from the critical point ($\lambda = 0.9$, left column), irrespective of the integration time (vertical axis), and the inputs can be perfectly assigned from the output.
When shifting $\lambda$ closer to the critical point, fluctuations increase such that with insufficient integration time, the errors appear but can be remedied when the integration time is increased.

To formalize this intuition, we consider the input-output distribution $P(o^T(t)|h)$ to observe an output $o^T(t)$ in response to a specific input $h$ at time $t$.
In the following, we will omit the time argument for brevity.
Now, if two inputs $h_1$ and $h_2$ were equally likely to be presented, then the overlap between $P(o^T|h_1)$ and $P(o^T|h_2)$ quantifies the minimal discrimination error~\cite{berens_neurometric_2009} of an ideal observer:
\begin{equation}
\label{Eq:MDE}
    \mathcal{E}(h_1,h_2)=\frac{1}{2}\int \min\left\{P(o^T|h_1),P(o^T|h_2)\right\}do^T.
\end{equation}
Computing this error for the stimuli in our example (Fig.~\ref{fig1}c), we find that $\mathcal{E}$ increases with $\lambda$ and decreases with $T$, which matches our just gained intuition.
In our example, variability in $o^T$ comes from observing stochastic, correlated dynamics ($\lambda$) with a finite integration time $T$ plus noise. However, our logic remains the same for other causes of variability.

As a next step, we define a set of \emph{discriminable inputs} that can be sufficiently well distinguished from observing only the output.
We call two inputs $\varepsilon$-discriminable if the overlap of the response distributions generated by the inputs is smaller than an error threshold $\varepsilon$.
Formally speaking, a set of $\varepsilon$-discriminable inputs $\mathcal{H}=\{h_1,h_2, \ldots, h_{n_d}\}$, with $h_0 = 0$ and $h_{n_d +1} = :h_\infty$, is a set for which $\mathcal{E}(h_i, h_j)\leq\varepsilon$ for all $i \neq j, \: i,j \in [0, n_d+1]$, where $h_\infty$ is an input that generated saturated output $\langle o^T\rangle = 1$.
Finding the maximal (in the sense of cardinality) set of discriminable inputs is a close-packing problem without a unique solution.
To circumvent this complication, we propose the following algorithm: start by finding $h_1^\mathrm{left} = \min\{h >h_0 = 0 : \mathcal{E}(h_0, h)\leq\varepsilon\}$, and then proceed by induction to $h^\mathrm{left}_{i+1} = \min \{h > h_i^\mathrm{left} : \mathcal{E}(h_i^\mathrm{left}, h)\leq\varepsilon \}$, (Fig.~\ref{fig1}, see Methods for more details).
We stop at the first $i$ such that $\mathcal{E}(h_{i+1}^\mathrm{left},h_\infty) > \varepsilon$ and get this way $n_d^{\mathrm{left}} = i$.
We repeat the same procedure starting from the right with $h_1^\mathrm{right} = \max\{h < h_{\infty} : \mathcal{E}(h, h_\infty)\leq\varepsilon \}$ and iterate until $\mathcal{E}(h_{i+1}^\mathrm{right},0) > \varepsilon$ to find $n_d^\mathrm{right}$.
Our final estimate of discriminable inputs cardinality is the average $n_d = 1/2 (n_d^{\mathrm{left}} + n_d^{\mathrm{right}})$.

While this algorithm is numerically straightforward, it comes with a technical challenge: it requires iterative evaluation of $P(o^T|h)$ for continuous values of $h$.
This is not a problem for our analytical solutions, but it becomes intractable for the actual numerical model because each $P(o^T|h)$ is a result of a long simulation.
To tackle this, we measure the distribution $P(a^T|h)$ of pure network activity $a^T(t)$ for a broad range of $h$, $T$, and $\lambda$ values, notice that they can be well approximated by a Beta distribution Beta($\alpha$,$\beta$), and train a neural network to learn the parameters $(\alpha,\beta)$ as a function of $(T, h, \lambda)$ to interpolate between them (see Methods and extended data figure Fig.~\ref{fig:extData_workflow}).

\begin{figure*}[t]
\centering
\includegraphics[]{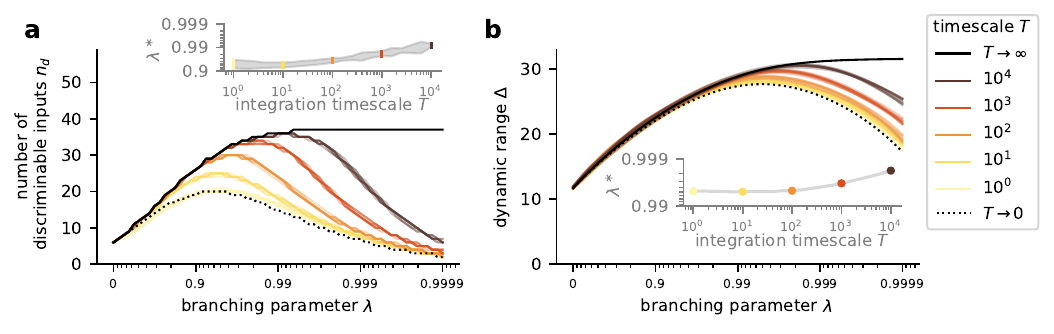}
\caption{%
\fs{Dynamical regime for optimal information transmission depends on the readout integration timescale.}
Discriminating input rates from a noisy output (cf.~\fig\ref{fig1}), we observe that \textbf{a)} the number of discriminable inputs as well as \textbf{b)} the dynamic range are maximal for sub-critical networks ($\lambda<1$).
With increasing timescale $T$, the maximum $\lambda^\ast$ moves closer to the critical point as demonstrated in the insets.
Parameters: $N=10^4$, $K=10^2$, $\mu=0.2$, $\nu=0.2$, $\sigma=10^{-2}$, $\varepsilon=10^{-1}$.
\label{fig2}
}
\end{figure*}

From the set of $\varepsilon$-discriminable inputs, we can now construct measures for information processing capabilities using finite integration time (Fig.~\ref{fig2}).
Let us start with the number of $\varepsilon$-discriminable inputs $n_d$ (Fig.~\ref{fig2}a).
Because the task requires sufficient coupling between the input and output population, $n_d$ is very small with $\lambda =0 $ and first increases with $\lambda$.
However, $n_d$ exhibits a maximum at a $T$-dependent subcritical $\lambda<1$, above which it decays for $\lambda\to 1$.
Our numerical results interpolate between the analytical predictions for $ T \to \infty$ (solid line) and $T \to 0$, indicating that every finite integration time will have an optimal $0<\lambda<1$, while for infinite integration time, $n_d$ is bound by the Gaussian noise of the readout.

Let us now turn to the dynamic range, which can be naturally generalized to account for fluctuations by choosing as bounds the first and last inputs that can be discriminated from the boundaries.
We thus define our \textit{dynamic range} as
\begin{equation}
    \Delta = \Delta(\varepsilon) = 10\log_{10}(h_1^\mathrm{right}/h_1^\mathrm{ left }),
\end{equation}
where the $h_1^\mathrm{ left }$ and $h_1^\mathrm{right}$ are the minimal and maximal inputs that can be discriminated from $h_0=0$ and $h_\infty$, respectively, with error not surpassing $\varepsilon$.
$\Delta$ depends on the specific choices of the discrimination error $\varepsilon$ and the variance of the Gaussian noise $\sigma$, which we can be tuned to recover the typical 10\%-90\% bounds of the established dynamic range~\cite{kinouchi_optimal_2006} for $T\to\infty$.
For finite $T$, our numerical estimates interpolate well between the analytical bounds (see SM for comparison with full readout).
Importantly, a finite $T$ results in a substantial reduction of the dynamic range in the vicinity of the critical point, i.e., for $\lambda\approx 1$, but only a slight reduction at small $\lambda$.
As a result, the dynamic range develops a $T$-dependent maximum, which is, however, different from the maximum of $n_d$ (insets in Fig.~\ref{fig2}).

Our results establish a connection between sensitivity (governed by the distance to criticality), confidence (capturing the probability of wrong classification), and integration time in a recurrent network of excitatory stochastic neurons.
While we primarily focused on discriminability as a function of the distance to criticality for a given integration time, we can also make statements about how $T$ and $\varepsilon$ affect the discriminability.
For any fixed $\lambda$ (vertical slice in \fig\ref{fig2}), we find that both measures of discriminability increase monotonically with $T$.
Also, by construction, the discriminability has to increase monotonically with the discrimination error $\varepsilon$.
Still, both $T$ and $\varepsilon$ affect the peaks in our measures of discriminability and thereby define the optimal state.
For finite $T$, the optimal balance is achieved by subcritical networks.

We expect that our insight about optimal sensitivity away from criticality will occur similarly in other stochastic systems, where emergent properties near criticality are beneficial for solving tasks in the presence of increasing stochastic fluctuations.
On which side of the transition this optimum lies will, however, depend on both the task and the type of phase transition, e.g., absorbing-to-active, transition to chaos, or a bifurcation -- see Ref.~\cite{munoz_colloquium_2018} for an overview.
For example, in the case of transition to chaos, it was shown that deviations toward the supercritical side, deeper into the chaotic regime, allow for slower integration times and are thus beneficial in the presence of noise~\cite{toyoizumi_beyond_2011}.
Additionally, emergent critical fluctuations do not necessarily have to align with the neural population activity; examples include a large dispersion of correlations~\cite{dahmen_second_2019} or low-dimensional subspaces~\cite{fontenele_low-dimensional_2024}.

Since near-critical dynamics imply a finite autocorrelation time~\cite{henkel_non-equilibrium_2008, tauber_critical_2014}, our results align well with the observation of finite timescales in neurophysiological data~\cite{uchida_seeing_2006, wilting_inferring_2018, zeraati_intrinsic_2023}.
While there is clear evidence for sensory integration, many perceptual tasks are solved in short times of less than a second~\cite{uchida_seeing_2006}.
This limited temporal integration can be due to non-stationary information rates, temporal correlations (such as in our recurrent dynamics), or leaky integrators (such as in our readout).
In our model, the recurrent autocorrelation timescale can be estimated assuming a linear autoregressive representation~\cite{wilting_inferring_2018, wilting_operating_2018}, yielding $\tau\approx-\Delta t/\ln(\lambda)$.
The confidence-dependent optima in Fig.~\ref{fig2} thus correspond to $\tau\approx\SI{10}{}-\SI{100}{ms}$, assuming a timestep $\Delta t=\SI{1}{ms}$, or to $\tau\approx\SI{50}{}-\SI{500}{ms}$ for a timestep $\Delta t=\SI{5}{ms}$ that could comprise various propagation delays and raise times.
This is consistent with empirical evidence of cortical timescales in the range of $\SI{50}{ms} - \SI{1}{s}$~\cite{murray_hierarchy_2014, rudelt_signatures_2024, shi_brain-wide_2025}.
Also, it is consistent with the recently observed adaptation to task requirements~\cite{zeraati_intrinsic_2023, cramer_control_2020}, which in our case would correspond to a change in $\varepsilon$ and $T$.

On the side of artificial networks, our results provide a new perspective on the reservoir-computing paradigm with typically memory-less readout signals~\cite{tanaka_recent_2019}.
While noise-free continuous formulations, like the echo-state network~\cite{jaeger_echo_2001}, allow to readout information about the past from standard nonlinear dynamic considerations, any system that comes with noise could benefit from integrating the readout over time, as was demonstrated recently for active particles~\cite{wang_harnessing_2024}.
In light of our results, an instantaneous readout would require a larger distance to criticality for optimal discriminability.
However, leaky readout units could allow the reservoir to be tuned closer to criticality, thereby benefiting from the edge-of-chaos sensitivity.

To summarize, given a readout with finite integration time, we find maximal discriminability for close-to-critical dynamics.
The intuitive reason for our finding is that emergent temporal fluctuations close to a critical phase transition can smear out the signal if they are aligned with the readout, and thereby hinder discrimination.
Since the network sensitivity is maximal at criticality, this implies a trade-off between sensitivity and discriminability.
Our results thereby add to the hypothesis that living systems need to adjust their state to optimally balance opposing demands depending on the specific processing tasks at hand~\cite{wilting_operating_2018, dahmen_strong_2022, khajehabdollahi_when_2022, zeraati_intrinsic_2023}.

\begingroup
\section*{Methods}
\paragraph*{Neural Network Model:}
We consider a network of $N=10^4$ binary spiking neurons, each described by a state variable $s_i$ that can be active ($s_i(t)=1$) or inactive ($s_i(t)=0$).
Time evolves in discrete steps of $\Delta t$.
Neurons can be activated by recurrent input from other neurons with probability $p^\mathrm{rec}[s_i(t+\Delta t)=1 | s(t)]=f(\sum_j w_{ij} s_j(t))$, where $w_{ij}$ are directed coupling weights (not symmetric) and $f(x)$ is a rectified linear function with $f(x)=0$ for $x<0$, $f(x)=x$ for $0<x<1$, and $f(x)=1$ for $x>1$.
The connectivity matrix $W = \left (w_{ij} \right)$ is a sparse matrix with mean degree $K=10^2$, where non-zero edges are selected with probability $K/N$ and diagonal entries are removed.
Non-zero weights are set to $w_{ij} = \lambda/K_i$, where $K_i$ is the indegree of neuron $i$ corresponding to the number of non-zero weights in row $i$.
Thereby, each neuron has the same maximal input of $\sum_j w_{ij}=\lambda$, and $\lambda$ is the largest eigenvalue of the connectivity matrix $W$.
Thereby, all neurons receive the same mean incoming weight $\sum_j w_{ij}=\lambda$, and $\lambda$ is the largest eigenvalue of the connectivity matrix $W$.

In addition to the recurrent activation, a random subset of $N^\mathrm{in} = \mu N$ neurons receive external input.
The external input is modeled as a Poisson process with rate $h$ or equivalently as an activation probability $p^\mathrm{ext}[s_i(t+ \Delta t)=1]=1-e^{-h\Delta t}$, which causes neurons to fire independently and irregularly.

The output is defined by an exponential smoothing of the spikes from a random subset of $N^\mathrm{out}=\nu N$ neurons plus Gaussian noise, cf. Eq.~\eqref{eq:output}.
Let us denote the pure network output as $a^T(t) = 1/N^\mathrm{out}\sum_{i\in N^\mathrm{out}}\sum_{\{k | \, t_i^k < t\}}e^{(t-t_i^k)/T}$, where the sum goes over all $i$ in the subset of output neurons and all spike times $k$ with $t_i^k<t$.
For discrete time steps, these exponential kernels can be implemented as standard exponential smoothening
\begin{equation}
  a^T(t) = (1-c^T)a^T (t-\Delta t) + c^T \frac{1}{N^\mathrm{out}}\sum_{i \in N^\mathrm{out}} s_{i } (t),
\end{equation}
with $c^T= 1-e^{-\Delta t/T}$.
Iterative substitution yields a geometric sequence as the discrete realization of the desired exponential function.

\begin{figure*}
    \centering
    \includegraphics[width=1\textwidth]{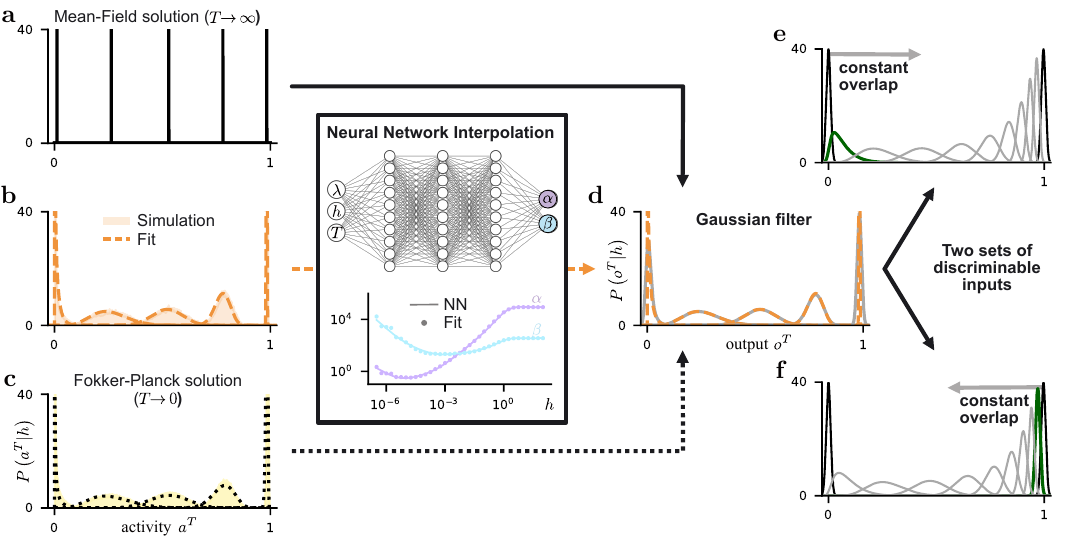}
    \caption{\fs{Workflow to calculate $\varepsilon$-discriminable inputs}.
    In the first step (a-c), we obtain the distribution $P(a_T|h)$ of pure output activity in a reservoir (with parameter $\lambda$) subject to an input $h$.
    (\textbf{a}) For $T\to\infty$, these distributions become $\delta$-distributions that we calculate using our mean-field approximation.
    (\textbf{b)} For finite $T$, we perform many numerical simulations and fit a Beta distribution to the data using maximum likelihood estimates (orange examples obtained for $T=100$).
    We then use these fit results to train a deep neural network as a general function approximation that interpolates the fit parameters $(\alpha,\beta)$ for all simulation parameters $(\lambda, h, T)$.
    (\textbf{c}) In the limit $T\to 0$, we obtain analytical results by solving the Fokker-Planck equation that is in good agreement with corresponding numerical data for $T=1$ (yellow histograms).
    (\textbf{d}) In the next step, we obtain the distribution of noisy output responses $P(o^T|h)$ by a convolution of $P(a^T|h)$ with a Gaussian $\mathcal{N}(0,\sigma^2)$ of small variance $\sigma^2$.
    This step allows i) to connect to previous mean-field results for $T\to\infty$~\cite{kinouchi_optimal_2006} and ii) circumvents numerical intricacies for finite $N$ at the boundaries.
    The example compares beta distributions from the neural network interpolations (orange) with their corresponding noisy distributions (gray), which mostly differ at the boundaries.
    (\textbf{e-f}) In the last step, we determine two sets of discriminable inputs that can be discriminated from reference distributions for vanishing input (left Gaussian distribution, black) and diverging input (right Gaussian distribution, black).
    For this, we start from the left and right references and perform iterative bisection searches in $h$ to find input values whose response distributions overlap exactly $\varepsilon$ with the previous one.
    The dynamic range is calculated from the smallest and largest inputs (marked green).
    The number of discriminable inputs is obtained as the average size of the sets.
    Examples are shown for $\lambda=0.999$, $\varepsilon=0.1$, $\sigma=0.01$.
    Example distributions for $h\in\left[5.6\cdot 10^{-5}, 1.8\cdot 10^{-3}, 5.6\cdot 10^{-3}, 1.8\cdot 10^{-2}, 3.2\cdot 10^{-1} \right]$
    }
    \label{fig:extData_workflow}
\end{figure*}
\paragraph*{Neural-Network approximation of output distribution:}
The stochastic simulations yield $P\left(a^T|h\right)$ by aggregating all measurements $a^T(t)$ after proper equilibration for specific values of $h$.
However, our estimation of discriminable inputs requires a distribution of $P\left(a^T|h\right)$ for any $h$ that can be achieved using interpolation.
To solve this, we first notice that $P\left(a^T|h\right)$ can be well approximated by a Beta distribution $\mathrm{Beta}(\alpha,\beta)$.
We obtain $(\alpha, \beta)$ as a maximum-likelihood estimate from simulations with parameters $\theta=(T, h, \lambda)$.
We scan the parameter space logarithmically in the ranges $1-\lambda\in\left[10^{-4}, 1\right]$, $h\in\left[10^{-6}, 10^2\right]$ and $T\in\left[1, 10^4\right]$ and train a dense 3-layer neural network to approximate the functions $\alpha(\theta)$ and $\beta(\theta)$, cf. Fig.~\ref{fig:extData_workflow}.
This exploits the fact that neural networks can act as general function approximators~\cite {hornik_multilayer_1989}.
Here, we choose 3 layers with 60 neurons each and a hyperbolic tangent ($\tanh$) activation function, following previous approaches to fitting scaling functions~\cite{dornheim_static_2019}.
We found good fits when scaling input and output parameters into the domain $[-1,1]$.
To ensure that the distribution mean increases monotonously with $h$ (relevant for the discriminable inputs), we further added a regularization term that penalizes deviations of the mean $\langle a^T\rangle = \alpha/(\alpha+\beta)$ from the mean-field solution Eq.~\eqref{eq_mean-field}, essentially implementing a physics-informed regularization.

\paragraph*{Mean-field solution for the limit \mbox{$T\to\infty$}:}
For simplicity, we perform mean-field computation for the case of the read-out population coinciding with the whole network.
For $T\to\infty$, we can neglect fluctuations such that the $P(a^T|h)$ becomes a delta distribution at the mean-field activity $a^\infty = a = \lim_{T\to \infty}  1/N \sum_{i=1}^N 1/T \sum_{t=1}^T  a_i(t) $.
To estimate the mean activity, we need to separate the network into the part that receives input with $N^\mathrm{in}=\mu N$ neurons and mean activity $a^\mathrm{in}$, and the rest of $N^\mathrm{rest}=(1-\mu)N$ neurons with mean activity $a^\mathrm{rest}$, such that the mean activity is $a = \mu a^\mathrm{in} + (1-\mu) a^\mathrm{rest}.$

Since each neuron is randomly connected to any other neuron in the network with the same total weight $\lambda=\sum_{ij}w_{ij}/N$, we can approximate the probability of recurrent activation
\begin{equation}\label{eq_mean-field_prob}
    \overline{p^\mathrm{rec}}=\overline{\sum_{j} w_{ij}s_j(t)}\approx\lambda\left[\mu a^\mathrm{in} +(1-\mu)a^\mathrm{rest})\right].
\end{equation}
Averaging out temporal fluctuations, we find that the mean activity equals the probability of activation.
For those neurons that can only be excited recurrently, we thus obtain $a^\mathrm{rest} = \overline{p^\mathrm{rec}}$.
For those neurons that receive external input, we need to take coalescence into account~\cite{zierenberg_description_2020} and find $a^\mathrm{in} = 1-(1-\overline{p^\mathrm{rec}})(1-p^\mathrm{ext})$.
This leaves us with a system of self-consistent equations
\begin{align}
    a^\mathrm{rest} &=\lambda\left[\mu a^\mathrm{in} + (1-\mu)a^\mathrm{rest}\right]\label{eq_mean-field_rest} \\
    a^\mathrm{in} &= 1-\left(1-\lambda\left[\mu a^\mathrm{in} + (1-\mu)a^\mathrm{rest}\right]\right)\left(1-p^\mathrm{ext}\right)\label{eq_mean-field_in}
\end{align}
that can be solved to yield
\begin{align}
    a=\frac{\mu p^\mathrm{ext}}{1-\lambda\left(1-\mu\right)-\lambda \mu \left(1-p^\mathrm{ext}\right)}\label{eq_mean-field}
\end{align}

\paragraph*{Mean-field solution for $T\to 0$:}
In the limit to continuous time, we can model the probability of neural activation and deactivation as a birth-death process with birth rate $\Omega_+(A)$ and death rate $\Omega_-(A)$, where $A$ is the number of active neurons.
The time evolution of the probability distribution $P(A,t)$ is then described by the master equation

\begin{align}
    \frac{d}{dt}P(A,t) =  &\Omega_+(A-1)~P(A-1,t) \\
     + &\Omega_-(A+1)~P(A+1,t) \\
     - &\left(\Omega_+(A) + \Omega_-(A)\right)~P(A,t)
\end{align}
Using a Kramers-Moyal expansion up to second order~\cite{risken_fokker-planck_2012}, we obtain the Fokker-Planck equation (see SM)

\begin{equation}
    \frac{d}{dt}P(A,t) = - \frac{d}{dA}\left[f(A)P(A,t)\right] + \frac{1}{2}\frac{d^2}{dA^2}\left[g(A)P(A,t)\right],\nonumber
\end{equation}
with a ``drift'' term $f(A)=\Omega_+(A) - \Omega_-(A)$ and a ``diffusion'' term $g(A)=\Omega_+(A) + \Omega_-(A)$.
The solution of the stationary Fokker Planck equation, $\frac{d}{dt}P(A,t)=0$, is

\begin{equation}
    P(A)\propto \frac{1}{g(A)} \exp\left\{2\int_0^A \frac{f(x)}{g(x)}dx\right\},
\end{equation}
which can solved numerically once birth and death rates are specified.

To specify the birth and death rates, we assume that inactive neurons can create activity by becoming active, while active neurons destroy activity by becoming inactive.
If $p^\mathrm{a}$ is the probability to activate any neuron in the next time step and there are $A$ out of $N$ neurons currently active, then we find
\begin{align}\nonumber
    \Omega_+(A) &= (N-A)~p^\mathrm{a}\\
    \Omega_-(A) &= A~(1-p^\mathrm{a})
\end{align}
Since the activation probability depends on whether the neuron receives external input or not, we need to distinguish between those neurons that receive input, $N^\mathrm{in}$, and those that can only be activated recurrently, $N^\mathrm{rest}$.
While for the latter, we can identify $p^\mathrm{a} = p^\mathrm{rec}$,  we need to account for coalescence in the former case and obtain $p^\mathrm{a} = 1- (1-p^\mathrm{rec})(1-p^\mathrm{ext})$.

To obtain an expression for $p^\mathrm{rec}$, we assume a mean-field setting where each connected neuron is described by its mean activity.
Then Eq.~\eqref{eq_mean-field_prob} yields the probability $\overline{p^\mathrm{rec}}(A^\mathrm{in}, A^\mathrm{rest})=\lambda\left[A^\mathrm{in} + A^\mathrm{rest}\right]/N$, which depends on the activity in both subsets and thereby couples their Fokker Planck equations.
To solve our Fokker Planck equations for $N^\mathrm{in}$ and $N^\mathrm{rest}$, we decouple this probability by replacing one variable via its mean-field equation as a function of the other.
Specifically, we start from Eq.~\eqref{eq_mean-field_in} to rewrite $A^\mathrm{in} = \mu\frac{N p^\mathrm{ext} + \lambda (1-p^\mathrm{ext}) A^\mathrm{rest}}{1-\mu\lambda(1-p^\mathrm{ext})}$ and get
\begin{equation}
    p^\mathrm{rec}(A^\mathrm{rest}) = \lambda\frac{A^\mathrm{rest}/N + \mu p^\mathrm{ext}}{1-\mu\lambda(1-p^\mathrm{ext})}.
\end{equation}
Similarly, we start from Eq.~\eqref{eq_mean-field_rest} to rewrite $A^\mathrm{rest} = \frac{(1-\mu)\lambda A^\mathrm{in}}{1-(1-\mu)\lambda}$ and get
\begin{equation}
    p^\mathrm{rec}(A^\mathrm{in}) = \lambda\frac{A^\mathrm{in}/N}{1-(1-\mu\lambda)}
\end{equation}
We can then independently solve the Fokker Planck equations for $P(A^\mathrm{rest})$ with
\begin{align}
    \Omega_+(A^\mathrm{rest}) &= (N-A^\mathrm{rest})~\lambda\frac{A^\mathrm{rest}/N + \mu p^\mathrm{ext}}{1-\mu\lambda(1-p^\mathrm{ext})} \\
    \Omega_-(A^\mathrm{rest}) &= A^\mathrm{rest}~\left(1-\lambda\frac{A^\mathrm{rest}/N + \mu p^\mathrm{ext}}{1-\mu\lambda(1-p^\mathrm{ext})}\right)
\end{align}
as well as $P(A^\mathrm{in})$ with
\begin{align}
    \Omega_+(A^\mathrm{in}) &= (N-A^\mathrm{in})~\lambda\frac{A^\mathrm{in}/N}{1-(1-\mu\lambda)}\\
    \Omega_-(A^\mathrm{in}) &= A^\mathrm{in}~\left(1-\lambda\frac{A^\mathrm{in}/N}{1-(1-\mu\lambda)}\right)
\end{align}
The solution for the total network activity $A=A^\mathrm{in} + A^\mathrm{rest}$ is obtained by the convolution $P(A)=P(A^\mathrm{in})*P(A^\mathrm{rest})$.

\endgroup

\section*{Data \& Code availability}
The simulation code, analysis pipeline, results, and scripts to produce the figures that support the findings of this study are available from GitHub at \href{https://github.com/sahelazizpour/Finite-Observation-Dynamic-Range}{sahelazizpour/Finite-Observation-Dynamic-Range}.

\section*{Acknowledgements}
J.Z. was supported by the Joachim Herz Stiftung.
J.Z. and V.P were funded by the Deutsche Forschungsgemeinschaft (DFG, German Research Foundation) - Project-ID 454648639 - SFB 1528 “Cognition of Interaction”.
A.L was supported by the Sofja Kovalevskaja Award from the Alexander von Humboldt Foundation.
All authors gratefully acknowledge support from the Max Planck Society.

\bibliography{sahel}

\clearpage
\renewcommand{\thefigure}{S\arabic{figure}}
\setcounter{figure}{0}
\renewcommand{\thetable}{S\arabic{table}}
\setcounter{table}{0}
\renewcommand{\theequation}{S\arabic{equation}}
\setcounter{equation}{0}

\onecolumngrid
\section*{Supplementary Information}

\begin{figure}
    \centering
    \includegraphics[]{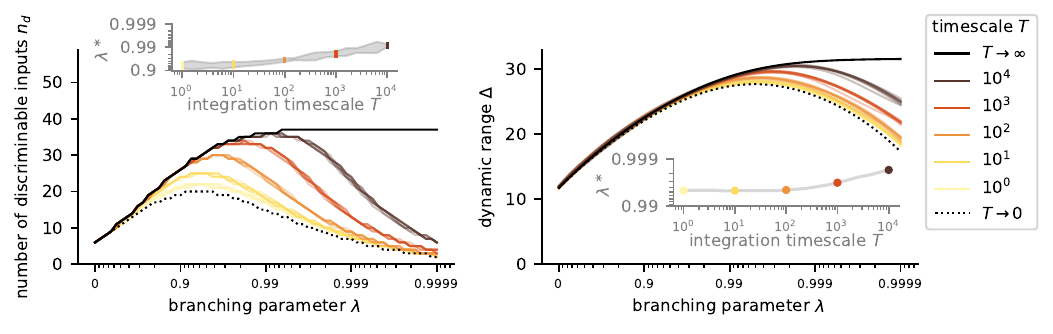}
    \caption{Number of discriminable input (left) and dynamic range (right) analogous to Fig.~2 main text but with $N^\text{out}=N$ to better compare to analytic solution.
    Notice that in this case the numerical results interpolate better between the analytic limits.%
    }
    \label{fig:results_nu=1}
\end{figure}

\subsection*{Results for full network readout}
For a fair comparison of our numerical solution with the analytical bounds $T\to 0$ and $T\to\infty$, which were derived for full network activity ($\nu=1$), we also measured the full network activity and applied the corresponding workflow to estimate our finite-integration-time measures (\fig\ref{fig:results_nu=1}).
As one can see, the numerical data lies perfectly within the analytical bounds besides minor deviations in the dynamic range for small $\lambda$.
These deviations can be attributed to the fluctuations in the random realizations of the networks that are neglected in the mean-field calculations and that are most relevant for weakly coupled systems with a lower probability of activity spread.

\subsection*{Recap of Kramers-Moyal expansion}

Let us consider the master equation of a birth-death process that we specified in our methods, with birth rate $\Omega_+(A)$ and death rate $\Omega_-(A)$,
\begin{equation}
    \frac{d}{dt} P(A,t) = \Omega_+(A-1)P(A-1,t) + \Omega_-(A+1)P(A+1,t) - (\Omega_+(A) + \Omega_-(A)) P(A,t). \label{Eq:Master_eq}
\end{equation}
This can be rewritten using a shift operator $E^l$ that acts on function $f(n)$ as $E^l[f(n)] = f(n+l)$, so that the master equation becomes
\begin{align}
    \frac{d}{dt} P(A,t) &= \sum_{l} (E^l-1)\Omega_{A \to A-l} P(A,t)\\
    &= (E^{-1}-1)\Omega_{A \to A+1} P(A-1,t) + (E^1-1)\Omega_{A \to A-1} P(A+1,t),  \label{Eq:Master_eq2}
\end{align}
where the last equality is a consequence of the incremental changes in steps of $\Delta A =1$, which we can assume for asynchronous updates, such that $E^{l}=0$ expect for $l=-1$ and $l=1$.
We can further substitute $\Omega_{A \to A+1} = \Omega_{+}$ and $\Omega_{A \to A-1} = \Omega_{-}$
For all sufficiently smooth functions $f$, we can write a Taylor expansion of $E^l[f(n)]$ that we truncate above second order, i.e.,
\begin{equation*}
    E^l[f(n)] = f(n+l) = f(n) + l \frac{d f(n)}{dn} + \frac{l^2}{2}\frac{d^2 f(n)}{dn^2}+ \ldots.
\end{equation*}
Thus
\begin{equation*}
    (E^l-1)[f(n)] = f(n+l) - f(n) = l \frac{d f(n)}{dn} + \frac{l^2}{2}\frac{d^2 f(n)}{dn^2}+ \ldots.
\end{equation*}
Inserting this truncated Taylor expansion into Eq.~\ref{Eq:Master_eq2}, we obtain a Fokker-Planck equation
\begin{equation*}
 \frac{d}{dt} P(A,t) = - \frac{d}{dA}(\Omega_{+}(A) P(A,t)) +  \frac{1}{2}\frac{d^2}{dA^2}(\Omega_{+}(A) P(A,t)) +  \frac{d}{dA}(\Omega_{-}(A) P(A,t)) +  \frac{1}{2}\frac{d^2}{dA^2}(\Omega_{-}(A) P(A,t)).
\end{equation*}
Rewriting this in the canonical form
\begin{equation}
   \frac{d}{dt} P(A,t) = - \frac{d}{dA} [f(A) P(A,t)] + \frac{1}{2}\frac{d^2}{dA^2}[g(A)P(A,t)], \label{Eq:Fokker-Planck}
\end{equation}
we can identify the drift term $f(A) = \Omega_{+}(A) - \Omega_{-}(A) $ and the diffusion term $g(A) = \Omega_{+}(A) + \Omega_{-}(A)$.

\end{document}